\documentclass[reprint,superscriptaddress,showpacs,amsmath,amssymb,twocolumn]{revtex4-1}

\usepackage{graphicx}
\usepackage{dcolumn}
\usepackage{bm}
\usepackage{verbatim}

\def\be{\begin{equation}}
\def\ee{\end{equation}}
\def\bea{\begin{eqnarray}}
\def\eea{\end{eqnarray}}

\begin{document}

\title{Transient photon echoes from donor-bound excitons in ZnO epitaxial layers}

\author{S.~V.~Poltavtsev}
\email{sergei.poltavtcev@tu-dortmund.de}
\affiliation{Experimentelle Physik 2, Technische Universit\"at Dortmund, 44221 Dortmund, Germany}
\affiliation{Spin Optics Laboratory, St.~Petersburg State University, 198504 St.~Petersburg, Russia}
\author{A.~N.~Kosarev}
\affiliation{Experimentelle Physik 2, Technische Universit\"at Dortmund, 44221 Dortmund, Germany}
\affiliation{Ioffe Physical-Technical Institute, Russian Academy of Sciences, 194021 St.~Petersburg, Russia}
\author{I.~A.~Akimov}
\affiliation{Experimentelle Physik 2, Technische Universit\"at Dortmund, 44221 Dortmund, Germany}
\affiliation{Ioffe Physical-Technical Institute, Russian Academy of Sciences, 194021 St.~Petersburg, Russia}
\author{D.~R.~Yakovlev}
\affiliation{Experimentelle Physik 2, Technische Universit\"at Dortmund, 44221 Dortmund, Germany}
\affiliation{Ioffe Physical-Technical Institute, Russian Academy of Sciences, 194021 St.~Petersburg, Russia}
\author{S.~Sadofev}
\affiliation{AG Photonik, Institut f\"ur Physik, Humboldt-Universit\"at zu Berlin, D-12489 Berlin, Germany}
\author{J.~Puls}
\affiliation{AG Photonik, Institut f\"ur Physik, Humboldt-Universit\"at zu Berlin, D-12489 Berlin, Germany}
\author{S.~P.~Hoffmann}
\affiliation{Department Physik \& CeOPP, Universit\"at Paderborn, 33098 Paderborn, Germany}
\author{M.~Albert}
\affiliation{Department Physik \& CeOPP, Universit\"at Paderborn, 33098 Paderborn, Germany}
\author{C.~Meier}
\affiliation{Department Physik \& CeOPP, Universit\"at Paderborn, 33098 Paderborn, Germany}
\author{T.~Meier}
\affiliation{Department Physik \& CeOPP, Universit\"at Paderborn, 33098 Paderborn, Germany}
\author{M.~Bayer}
\affiliation{Experimentelle Physik 2, Technische Universit\"at Dortmund, 44221 Dortmund, Germany}
\affiliation{Ioffe Physical-Technical Institute, Russian Academy of Sciences, 194021 St.~Petersburg, Russia}

\date{\today}

\begin{abstract}
The coherent optical response from 140~nm and 65~nm thick ZnO epitaxial layers is studied using transient four-wave-mixing spectroscopy with picosecond temporal resolution. Resonant excitation of neutral donor-bound excitons results in two-pulse and three-pulse photon echoes. For the donor-bound A exciton (D$^0$X$_\text{A}$) at temperature of 1.8~K we evaluate optical coherence times $T_2=33-50$~ps corresponding to homogeneous linewidths of $13-19~\mu$eV, about two orders of magnitude smaller as compared with the inhomogeneous broadening of the optical transitions. The coherent dynamics is determined mainly by the population decay with time $T_1=30-40$~ps, while pure dephasing is negligible in the studied high quality samples even for strong optical excitation. Temperature increase leads to a significant shortening of $T_2$ due to interaction with acoustic phonons. In contrast, the loss of coherence of the donor-bound B exciton (D$^0$X$_\text{B}$) is significantly faster ($T_2=3.6$~ps) and governed by pure dephasing processes.
\end{abstract}

\maketitle

The optical properties of ZnO are of great interest for applications in ultraviolet (UV) light emitting devices \cite{TsukazakiNature2005,RyuAPL2006}, polariton lasers \cite{PolLaserPRL2014}, UV-sensitive photodetectors \cite{SharmaAPL2002} and other photonic devices \cite{KlingshirnReview2007}. The main feature of this material is the large exciton binding energy of $\sim60$~meV \cite{Thomas1960}, so that exciton emission occurs even at room temperature \cite{RodinaPSS2004,HennebergerAPL2006}. Another remarkable property is the large exciton oscillator strength resulting in short optical lifetimes, which could be useful in applications requiring fast coherent control.

So far, the coherent optical properties of ZnO were studied on free excitons with short coherence times up to few ps \cite{ZhangAPL1999,HazuJAP2004,ShubinaPRB2013,Popov2014}. Optical excitation with spectrally broad femtosecond pulses was implemented and, consequently, quantum beats between various exciton states as well as strong many body interactions were demonstrated \cite{HazuJAP2004, ShubinaPRB2013}. For exciton complexes bound to impurities, e.g. donor-bound excitons, long-lived coherence is expected \cite{NollPRL1990}. In order to study distinct states exhibiting long coherence times, four-wave-mixing (FWM) spectroscopy is preferentially performed with resonant excitation of the exciton complex of interest using spectrally narrow pulses, even though the time resolution is reduced thereby. Furthermore, to overcome the large inhomogeneity of exciton transitions, which represents another intrinsic property of ZnO leading to sub-ps decay of macropscopic polarization, photon echo-based techniques are the best choice for studying the coherent dynamics of excitons \cite{KochPRB1993}.

In this letter, we demonstrate that for resonant optical excitation of the donor-bound A and B excitons in ZnO epitaxial layers with ps pulses the coherent response is given by photon echoes. The decay of the photon echo signals allows one to determine intrinsic properties of the single donor bound excitons, such as the coherence time $T_2$ and population decay time $T_1$. At temperature of 1.8~K, the coherence time of the A bound exciton (${\rm D^0X_A}$) is in the range of several tens of ps, corresponding to a homogeneous linewidth of about 10-20~$\mu$eV, which is two orders of magnitude smaller than the inhomogeneous broadening of D$^0$X$_\text{A}$. Our findings demonstrate that the ${\rm D^0X_A}$ in ZnO represents a promising two-level system which may be used for ultrafast optical control. In contrast, the decoherence of ${\rm D^0X_B}$ occurs one order of magnitude faster, while the population decay is approximately the same. In addition, we show that a temperature increase leads to significant shortening of the coherence times due to interactions with phonons.

{\it Samples and method.} For our study of localized exciton states in bulk ZnO, we used two ZnO epilayers grown by plasma-assisted molecular-beam epitaxy on c-plane (0001) sapphire substrates. Sample~I fabricated in Berlin (ZMO1031) is a 140~nm thick ZnO epitaxial layer surrounded by Zn$_{0.9}$Mg$_{0.1}$O layers with thicknesses of 100~nm and 1~$\mu$m from top and bottom, respectively. Sample~II fabricated in Paderborn (ZnO-385) is a 65~nm thick ZnO layer separated from the substrate by a 45 nm thick buffer of low-temperature grown ZnO. Both samples are deposited with a 1-2 nm-thick MgO nucleation layer.

\begin{figure}[h]
	\vspace{5mm}
	\includegraphics[width=\linewidth]{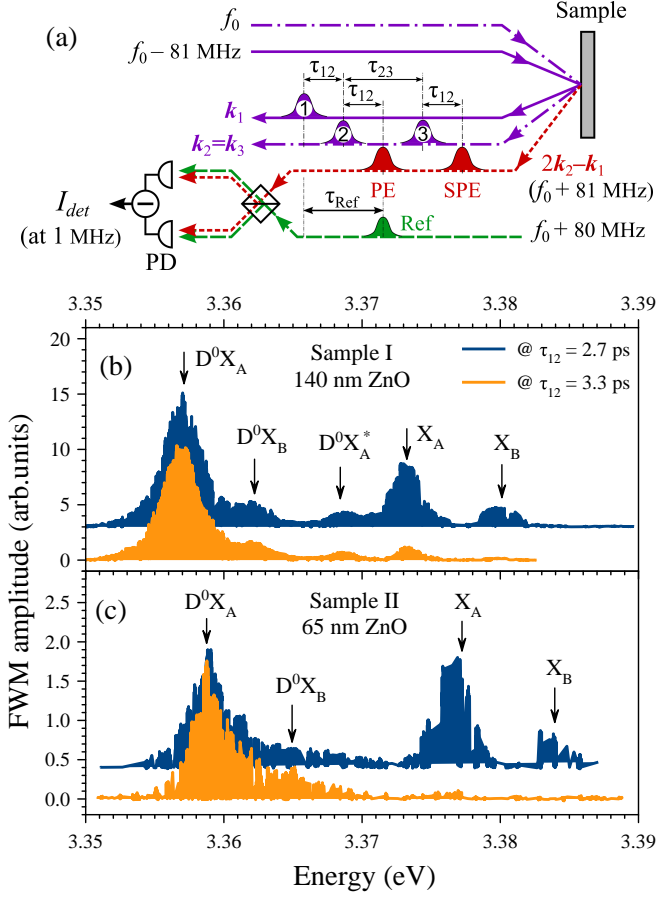}
	\caption{(Color) Transient FWM spectroscopy in ZnO: (a) Schematic of heterodyne time-resolved FWM experiment. $f_0$ is optical frequency of laser pulse. PD denotes photodetector. (b) and (c) FWM spectra  of samples~I and II, respectively, measured at second pulse delays $\tau_{12}=2.7$ ps and 3.3~ps. FWM detection was performed at $2\tau_{12}$. Arrows indicate transitions for X$_\text{A}$, X$_\text{B}$, D$^0$X$_\text{A}$, D$^0$X$_\text{B}$, and D$^0$X$_\text{A}^*$. $T=1.8$~K.}
	\label{setup}
\end{figure}

The coherent optical response was measured using a three-pulse degenerate FWM setup in reflection geometry. The samples were inserted in a helium bath cryostat and cooled down to 1.8~K. As laser source we used a Ti:Sapphire laser Mira-900 with pulse repetition rate of 75.75~MHz combined with an external second-harmonic generation unit which delivered frequency-doubled pulses in ultraviolet spectral range with 1.3~ps-duration. Optical excitation of the samples was done either by two pulses separated by time $\tau_{12}$ in the two-pulse photon echo (PE) experiment or by three pulses (the second and third pulses separated by $\tau_{23}$) in the stimulated photon echo (SPE) experiment \cite{BermanMalinovskyBook}, as schematically shown in Fig.~\ref{setup}(a). The laser photon energy was tuned in the spectral range of the free and donor-bound excitons in ZnO: $3.35-3.39$~eV. The first ($\bm{k}_1$) and second ($\bm{k}_2$) pulses hit the sample under angles of $\sim3^\circ$ and $4^\circ$, respectively, in a spot of about 250 $\mu$m diameter. The third pulse was collinear with the second one, $\bm{k}_3=\bm{k}_2$. The intensity of the excitation pulses was kept below 2~mW per beam ensuring the coherent response to be in the $\chi^{(3)}$ regime. The FWM signal was detected along the $\bm{k}_{\rm FWM}=2\bm{k}_2-\bm{k}_1$ direction. To detect the weak FWM signal, optical heterodyning and interference with a reference laser pulse were exploited~\cite{Hofmann-APL1996, LangerPRL2012}. Optical heterodyning was accomplished with two acousto-optical modulators (AOMs) acting as optical frequency shifters. The optical frequency of the first pulse was shifted to $f_1=f_0-81$MHz with the first AOM, while the optical frequency of the reference pulse separated by $\tau_{\rm Ref}$ from the first pulse was shifted to $f_{\rm Ref}=f_0+80$MHz with the second AOM. Thus, the optical frequency of the FWM signal is given by $f_{\rm FWM}=2f_0-f_1=f_0+81$MHz. The FWM and the reference beams were mixed in a silicon photodetector and the modulus of the cross-correlation of the FWM optical field with the reference pulse field, $I_{det}$, was detected at 1~MHz frequency by a fast lock-in amplifier. Additionally, the first beam was modulated by an optical chopper at 1~kHz frequency, at which synchronous detection by a slow lock-in amplifier was done. All beams were linearly co-polarized.

{\it Experimental results.} Figures~\ref{setup}(b) and \ref{setup}(c) show FWM spectra measured at $\tau_{\rm Ref}=2\tau_{12}$ on both samples for short $\tau_{12}$ delays of 2.7~ps and 3.3~ps. Four main transitions are seen in the spectra associated with the free A~exciton (X$_\text{A}$), free B~exciton (X$_\text{B}$), neutral donor-bound A~exciton (D$^0$X$_\text{A}$), and neutral donor-bound B~exciton (D$^0$X$_\text{B}$), in good correspondence with the absorption spectrum measured on sample~I \cite{PulsPRB2016}. Additional transition is seen for sample~I at $\sim3.369$~eV (about 11.7~meV above D$^0$X$_\text{A}$), as previously observed in photoluminescence excitation spectra and attributed to an excited $d$-state of the donor-bound exciton A (D$^0$X$_\text{A}^*$)~\cite{PulsPRB2016,LagardePRB2008,Meyer_PRB2010}. The exciton resonances in the 65~nm ZnO layer are shifted by $2.0-3.7$~meV relative to those in the 140~nm layer, which we attribute to presence of residual strain in the samples \cite{Wagner_PRB2013}. The free A and B exciton signals decay extremely fast on sub-ps timescale, so that they are already significantly damped in FWM when measured at $\tau_{12}=3.3$~ps. We estimate the coherence times of these free excitons to be below 1~ps, in line with previous studies~\cite{ZhangAPL1999, HazuJAP2004, ShubinaPRB2013}. Donor-bound excitons decay significantly slower so that they can be studied in detail with ps transient photon echo spectroscopy.

\begin{figure}[t]
	\vspace{5mm}
	\includegraphics[width=\linewidth]{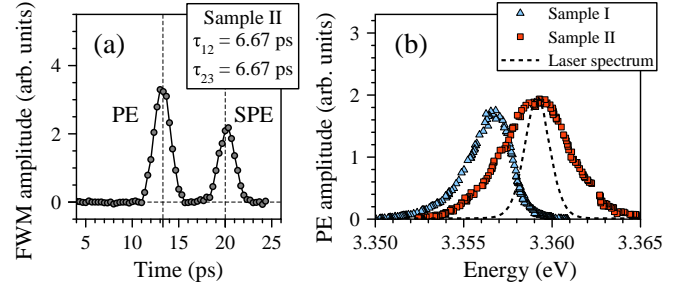}
	\caption{(Color)  Photon echo measurements on D$^0$X$_\text{A}$: (a) FWM transient measured in three-pulse configuration on sample~II at $\tau_{12}=\tau_{23}=6.67$~ps. (b)~PE spectra measured at $\tau_{12}=26.7$~ps and $\tau_{12}=13.3$~ps for samples I and II, respectively. Spectrum of ps-pulses is shown with dashed line.}
	\label{transients}
\end{figure}

FWM signal from D$^0$X$_\text{A}$ in form of photon echoes was observed on both samples. Figure~\ref{transients}(a) displays the FWM amplitude transient measured on sample~II in the three-pulse echo experiment at $\tau_{12}=\tau_{23}=6.67$~ps. This transient is composed of two echo pulses located at the times of the PE ($2\tau_{12}=13.3$~ps) and the SPE ($2\tau_{12}+\tau_{23}=20$~ps). The temporal profiles of PE and SPE are well fitted with Gaussians with a full width at half maximum (FWHM) of 2.1~ps. In this case the excited ensemble is defined by the laser spectrum, i.e.,  the latter is narrower than the inhomogeneous width of the optical transitions. Figure~\ref{transients}(b) shows PE spectra measured at $\tau_{12}=26.7$~ps for sample~I and 13.3~ps for sample~II. The spectral resolution in these measurements is given by the 1.7~meV FWHM of the laser spectrum, shown by the dashed line. Accordingly in sample I the inhomogeneous broadening of optical transitions is smaller, which we attribute to weaker strain gradients across the ZnO epilyaer due to presence of the intermediate ZnMgO buffer layer.

\begin{figure}[t]
	\vspace{5mm}
	\includegraphics[width=\linewidth]{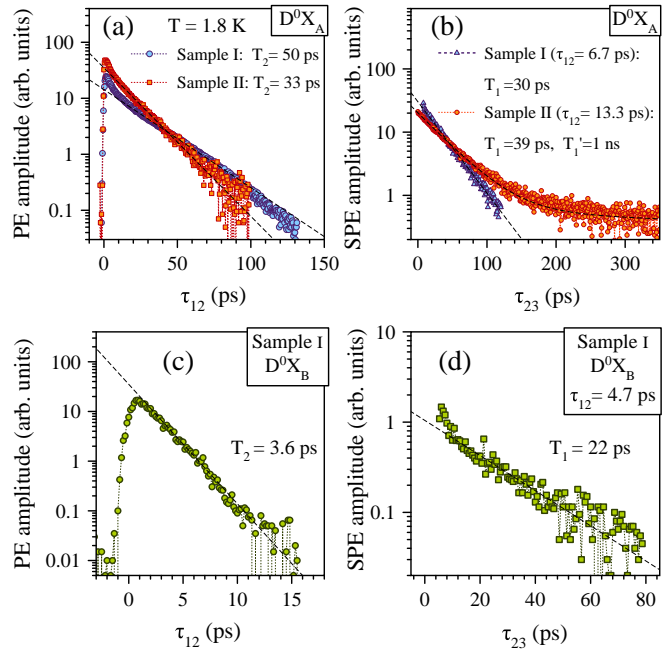}
	\caption{(Color) PE and SPE amplitude decays measured at $T=1.8$ K as function of delays $\tau_{12}$ and $\tau_{23}$, respectively. (a) PE decays and (b) SPE decays measured for D$^0$X$_\text{A}$ in samples I and II. (c) PE decay and (d) SPE decay measured from D$^0$X$_\text{B}$ (3.3627~eV) in sample~I. $\tau_{12}$ delays used and decay times extracted from exponential fits are given in panels.}
	\label{PE_SPE}
\end{figure}

To measure the $T_2$ and $T_1$ times of the donor bound excitons, we either vary the $\tau_{12}$ delay with PE amplitude detection or the $\tau_{23}$ delay with SPE amplitude detection. These data are summarized in Fig.~\ref{PE_SPE}. D$^0$X$_\text{A}$ PE decays measured on both samples in the two-pulse echo experiment are shown in Fig.~\ref{PE_SPE}(a). Both curves can be well described by mono-exponential decays, from which coherence times of $T_2=50\pm0.5$~ps for sample~I and $33\pm0.5$~ps for sample~II can be extracted. These correspond to homogeneous linewidths $\gamma_{\rm D^0X_A}=\hbar/T_2=13-19~\mu$eV, respectively. We emphasize that these values are more than two orders of magnitude smaller than the inhomogeneous widths of the optical transitions ($\sim1$~meV). The small difference in decoherence rates in the two samples can originate from different levels of impurities and defects. Nevertheless, the small homogeneous widths of the D$^0$X$_\text{A}$ optical transitions indicate a high crystal quality.

Figure~\ref{PE_SPE}(b) demonstrates the SPE decays measured at $\tau_{12}=6.67$~ps and 13.3~ps for samples I and II, respectively. The extracted decay times are $T_1=30\pm0.5$~ps for sample~I and $T_1=39\pm1$~ps for sample~II. Sample~II exhibits additionally a long-lived component with small amplitude decaying on a $1000\pm200$~ps timescale, which can be explained as follows: The first and second laser pulses create a spectral exciton population grating. During exciton decay, non-radiative relaxation processes can partially empty the excited state, leaving the ground state uncompensated even after exciton recombination \cite{Morsink1979,LangerNP2014}. The uncompensated spectral grating in the ground state can contribute to the long-lived echo component. This observation hints at a larger concentration of defects, e.g. deep impurities, in sample II.

For the D$^0$X$_\text{A}$ transition in sample I $T_2\approx 2T_1$ and, therefore, we conclude that pure dephasing processes, i.e. elastic scattering of excitons, can be neglected. Thus, the loss of coherence is attributed to the exciton population dynamics, namely energy relaxation including also radiative decay.  Interestingly, the value of $T_1=30$~ps is about 4-5 times shorter than the lifetime $\tau_0\approx140$~ps measured by time-resolved photoluminescence using a streak-camera on the same sample \cite{PulsPRB2016} and the $\tau_0\approx160$~ps for a similar ZnO epilayer \cite{LagardePRB2008}. This difference indicates that the SPE decay time $T_1$ observed here is limited not only by the exciton lifetime, but also by other energy relaxation processes, to which the photoluminescence is insensitive. The relaxation mechanisms behind this exciton decay shortening require further studies.

Next, we compare the coherence of ${\rm D^0X_A}$ and ${\rm D^0X_B}$ in sample I. Figures~\ref{PE_SPE}(c) and \ref{PE_SPE}(d) show the PE and SPE decays at the D$^0$X$_\text{B}$ optical transition, respectively. From exponential fits we evaluate a coherence time of $T_2\approx3.6$~ps and a population decay time $T_1\approx22$~ps for ${\rm D^0X_B}$. Thus, the homogeneous linewidth of the optical transition, $\gamma_{\rm D^0X_B}=180~\mu$eV, is about an order of magnitude larger than $\gamma_{\rm D^0X_A}$. The fast decoherence of ${\rm D^0X_B}$ cannot be attributed to energy relaxation because the population decay remains approximately the same as for ${\rm D^0X_A}$. Therefore, in contrast to ${\rm D^0X_A}$, pure dephasing dominates for ${\rm D^0X_B}$. This is a surprising result because the A and B donor bound excitons have similar binding energies and, therefore, occupy the same localization volume. A possible explanation for this unexpected behavior is that the ${\rm D^0X_B}$ states are located in close proximity to ${\rm D^0X_A}$ excited states \cite{Meyer_PRB2010}. Coupling of these states could explain the faster loss of coherence of ${\rm D^0X_B}$.

\begin{figure}[t]
	\vspace{5mm}
	\includegraphics[width=\linewidth]{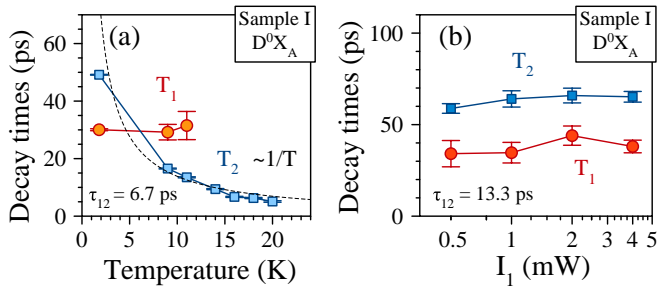}
	\caption{(Color) Results of PE and SPE decay measurements at D$^0$X$_\text{A}$ transition in sample~I. (a) Temperature dependence of $T_1$ and $T_2$ times. Dashed line is fit to $T_2(T)$ data with $\sim1/T$. (b) Dependence of $T_1$ and $T_2$ on first pulse intensity $I_1$. Intensities of second and third pulses are $I_2, I_3 < 0.6$~mW.}
	\label{PE_SPE_temperature}
\end{figure}

Finally, to get deeper insight into the coherence properties of donor-bound excitons, we study the PE decay rate in sample~I as function of temperature and optical excitation intensity. The temperature dependences of coherence and population decay times measured on D$^0$X$_\text{A}$ are shown in Fig.~\ref{PE_SPE_temperature}(a). While $T_1$ is temperature independent up to 12~K, the coherence time decreases as $\sim1/T$ in accord with the linear increase of the acoustic phonon population leading to pure dephasing of the excitons. As already mentioned before, at $T=1.8$~K, $T_2\approx2T_1$ indicating that additional irreversible dephasing mechanisms are negligible. It is also striking that both times, $T_2$ and $T_1$, are independent of optical excitation intensity as demonstrated in Figure~\ref{PE_SPE_temperature}(b) for variation of the first pulse intensity over almost one order of magnitude. We also checked that these times remain the same even when the second and third pulse powers are increased up to several mW.

{\it Conclusions.} Epitaxial ZnO layers were studied using coherent optical spectroscopy. Two-pulse and three-pulse photon echoes were observed from the donor-bound A and B excitons. ${\rm D^0X_A}$ shows a coherent dynamics on timescales of several tens of picoseconds, corresponding to a homogeneous linewidth of 13-19~$\mu$eV at $T=1.8$~K. That the D$^0$X$_\text{A}$ coherence time is independent of excitation intensity supports the hypothesis that these excitons can be coherently driven in a robust manner by the optical field. On the contrary, the D$^0$X$_\text{B}$ dephasing is much faster than that for D$^0$X$_\text{A}$ ($T_2=3.6$~ps and 50~ps, respectively), whereas the population decay times are comparable ($T_1=33$~ps for D$^0$X$_\text{A}$ and 22~ps for D$^0$X$_\text{B}$). This indicates that pure dephasing dominates for ${\rm D^0X_B}$, in contrast to ${\rm D^0X_A}$.
We also show that acoustic phonons play important role, limiting the donor-bound exciton coherence in ZnO at elevated temperatures. In comparison with other wide-bandgap wurtzite semiconductors, CdS was reported to show a coherence time of 800~ps, comparable with its lifetime of 1000~ps, for the neutral acceptor-bound exciton \cite{HoffmannPSS1998}. Due to the much larger oscillator strength, the coherent optical dynamics of the donor-bound exciton in ZnO is significantly shorter lived, making ZnO an attractive candidate for fast coherent control as compared to other wide-bandgap semiconductors.

{\it Acknowledgments.} We are grateful to A.V. Rodina for useful discussions. We acknowledge the financial support of the Deutsche Forschungsgemeinschaft through the Collaborative Research Centre TRR 142. S.V.P. thanks the Russian Foundation of Basic Research for partial financial support (contract no. 15-52-12016 NNIO$\_$a). M.B. acknowledges the partial financial support from the Russian Ministry of Science and Education (contract no. 14.Z50.31.0021).


%

\end{document}